\documentclass[12pt,a4paper]{article} 
\usepackage[pdftex,plainpages=false,pdfpagelabels]{hyperref}
\usepackage{calc,amsmath,amsthm,amssymb,color,graphics,natbib}
\usepackage{appendix}
\usepackage{fullpage}
\usepackage{fancyhdr}
\usepackage{standardmacros}
\usepackage{rotating}
\usepackage{lineno}

\newdimen\np
\title{MMANOVA: A general multilevel framework for multivariate analysis of
variance}
\author{
\begin{minipage}{0.3\textwidth}
\begin{center}
Steven Geinitz \\
\begin{normalsize}
\textit{geinitz@math.uzh.ch} \\
University of Zurich \\~
\end{normalsize}
\end{center}
\end{minipage}
\begin{minipage}{0.3\textwidth}
\begin{center}
Reinhard Furrer \\
\begin{normalsize}
\textit{furrer@math.uzh.ch} \\
University of Zurich \\~
\end{normalsize}
\end{center}
\end{minipage}
\begin{minipage}{0.3\textwidth}
\begin{center}
Stephan R. Sain \\
\begin{normalsize}
\textit{ssain@ucar.edu} \\
National Center for\\
\vspace{-1.5mm}
Atmospheric Research\\
\vspace{1.5mm}
\end{normalsize}
\end{center}
\end{minipage}
}
\date{}
\begin{document}
\maketitle
\begin{abstract}  
\vspace{-1cm}
\noindent
\linenumbers
\textbf{Abstract}
Classical analysis of variance requires that model terms be labeled as fixed or
random and typically culminate by comparing variability from each batch (factor)
to variability from errors; without a standard methodology to assess the
magnitude of a batch's variability, to compare variability between batches, nor
to consider the uncertainty in this assessment.  In this paper we support recent
work, placing ANOVA into a general multilevel framework, then refine this
through batch level model specifications, and develop it further by extension to
the multivariate case.  Adopting a Bayesian multilevel model parametrization,
with improper batch level prior densities, we derive a method that facilitates
comparison across all sources of variability.  Whereas classical multivariate
ANOVA often utilizes a single covariance criterion, e.g.\ determinant for Wilks'
lambda distribution, the method allows arbitrary covariance criteria to be
employed.  The proposed method also addresses computation.  By introducing
implicit batch level constraints, which yield improper priors, the full
posterior is efficiently factored, thus alleviating computational demands.  For
a large class of models, the partitioning mitigates, or even obviates the need
for methods such as MCMC.  The method is illustrated with simulated examples and
an application focusing on climate projections with global climate models. \\ \\
\noindent
\textit{Keywords}:\, Bayesian inference; Constraints; Mixed model; Variance components
\end{abstract}

\linenumbers
\section{Introduction}
Identifying and comparing variability among several factors is a fundamental
task of statistical analysis.  From initial exploratory steps, to model testing,
analysis of variance plays a vital role in the practice of statistics.
\citet{Gelm:05a} has outlined a general ANOVA methodology that fits a wide range
of models and summarizes results in a manner that facilitates interpretation
across different sources of variation.  \citet{Gelm:06} elaborate further,
describing the methodology in terms of a multilevel model.  One important
contribution of this approach is in providing summaries that are more
constructive than conclusions based on hypothesis tests.  This framework is
seeing usage by researchers from diverse fields e.g.\ ecology
\citep{Qian:Shen:07}, genetics \citep{Lein:08}, and climate
\citep{Sain:etal:11}.  

This paper presents a method that adopts the multilevel approach towards ANOVA,
then extends it to multivariate settings, yielding a multilevel multivariate
analysis of variance (MMANOVA) methodology.  The strategy of initially treating
all sources of variability in similar regard is naturally a part of the new
method.  We address known issues of applying constraints in variance analyses
\citep{Neld:77,Neld:94,Neld:99} by constraining batch levels so that prior
distributions are improper.  The extension is seen as a valuable contribution,
since multivariate ANOVA can further obfuscate model specification,
interpretation, and computation.  \citet{Kauf:Sain:10} offer an analysis of
variance procedure that aids in model specification and interpretation, but
requires computationally demanding MCMC steps.  Recent methods involving
approximations, e.g.\ integrated nested Laplace approximations (INLA)
\citep{Lind:etal:11,Rue:etal:09}, have allowed for MCMC to be eliminated in many
cases, significantly reducing computational demands.  While the range of
problems for which such methods are applicable is wide, the focus is not
typically on variance parameters.  Thus, the contribution of our work is in
promoting a general ANOVA methodology.  We accomplish this by supporting recent
work with the same goal, by refining the model specification, and by extending
this to multivariate cases.

In Section~\ref{sec:anova} we review the ANOVA formulation of \citet{Gelm:05a}
and summarize some details of the approach.  We then formally extend the idea to
the multivariate case, discussing technical and computational aspects.
Section~\ref{sec:examples} provides a demonstration of the method through a
simulation example and through a climatological application using data from
future climate projections given by several atmosphere-ocean general circulation
models and global emissions scenarios.  In Section~\ref{sec:discuss} we discuss
extensions and further computational benefits that are possible when the method
is applied to common  high-dimensional problems.

\section{Analysis of Variance}\label{sec:anova}
Analysis of variance is widely accepted as a means of partitioning variability
in a manner which allows it to be attributed to various factors.  An important
initial step in the analysis is considering each factor of the model to be fixed
or random.  This step, necessary in the classical setting, raises enough issues
that statisticians have been obliged to address the ``mixed models controversy''
\citep{Lenc:Sing:05, Voss:99}.  One might conclude that a consensus has still
not been reached, given that John Nelder deemed it necessary to reiterate the
requisite points of constraints and marginality over such a long period of time,
beginning with \citet{Neld:77} and most recently \citet{Neld:08}. 

The rest of the section outlines a recent attempt by \citet{Gelm:05a} towards a
more universal ANOVA methodology, then refines and extends the approach to a
multivariate context. 

\subsection{Multilevel ANOVA} \label{subsec:univaranova}
A fundamental contribution of the hierarchical regression approach to ANOVA
employed by \citet{Gelm:05a} has been to indiscriminately consider all
components in a model as random, thereby facilitating comparison across all
sources of variability.  The terminology is useful in supporting the
indiscriminate nature of the method.  The word \emph{batch} is applied to all
terms in the model, e.g.\ overall mean, factors, nested terms, interactions,
etc.  The nature of the variability from a batch is further distinguished.  The
distinction traditionally made by random and fixed effects is instead addressed
by considering a batch's \emph{super} and \emph{finite} population variance.  We
now summarize the recent shift in ANOVA as a methodology in terms of a
univariate linear model.

\subsubsection{Model Parametrization}\label{subsubsec:gelmananova} 
Following the notation of \citet{Gelm:05a}, observations $Y_i, i = 1, \dots, n$
are stated in terms of the additive decomposition

\vspace{-0.55cm}
\begin{align}
  Y_{i} & = \sum_{b=0}^{B} \beta^{(b)}_{j_{i}^{b}} 
    \label{eq:anovamodel1}, 
\end{align}
or the alternative regression formulation

\vspace{-0.55cm}
\begin{align}
  Y_{i} & = \sum_{b=0}^{B}\sum_{j=1}^{n_{B}} x_{i,j}^{(b)}\beta^{(b)}_{j^{b}}
    \label{eq:anovamodel2},
\end{align}
with $\beta^{(b)}_{j^{b}}$ denoting individual batch levels and $x_{i,j}$
denoting explanatory variables.  The regression formulation could be used for
the additive decomposition, with explanatory variables set to either $0$ or $1$.
Batch indices $b=0$ and $b=B$ will often correspond to an overall mean, $\mu$,
and to measurement errors, $\epsilon_i$, respectively, so that $n_0 = 1$, $n_B =
n$.  An individual batch is referenced by $b = 0, \dots, B$ and consists of
$n_{b}$ levels.  Individual levels of a batch are denoted as $\beta_{1}^{(b)},
\dots, \beta_{n_{b}}^{(b)}$ with $j_1^b, \dots, j_n^b$ replicating the levels so
that each observation is associated with exactly one batch level.  We
acknowledge that the additional level of subscripts and superscripts may seem
contrived to some, although it is necessary for the general case.  In practice
the number of batches in the model is reasonable so that this can be avoided, as
done in Section~\ref{sec:examples}.   Additional sub or superscripts can often
be dropped.  For example, a batch in its entirety is denoted by  $\beta^{(b)} =
\{ \beta^{(b)}_1, \dots, \beta^{(b)}_{n_b} \}$.  Given a batch, and an assumed
distribution on model errors, a conventional fixed effects analysis often
corresponds to the test $H_{0}\!: \beta^{(b)}_{j}\!=\!0$ for $j = 1, \dots,
n_b$.  While for a random effects model, assuming the $n_{b}$ levels of each
batch $b$ to be modeled as Gaussian

\vspace{-0.55cm}
\begin{align} 
  \beta_{j}^{(b)} \mid \sigma^2_b & \sim \mathrm{N}(0, \; \sigma^{2}_{b}),\qquad
  j=1,\dots,n_{b}, \label{eq:randeffeq}
\end{align}
a test for significant batch variation would be $H_{0}\!: \sigma^2_{b} = 0$.
Alternatively, the proposed methodology identifies two representations of
variation of a given batch.  The superpopulation variance, $\sigma^{2}_{b}$,
corresponds to the variance of all potential, possibly infinitely many, levels
of a batch.  The finite-population variance represents variability of the
specific set of batch levels that have been realized.  Super and
finite-population variances can be roughly related to the random effect variance
component estimate, and the fixed effect within-group sum of squares,
respectively.  As an example, consider batch $b$ and its vector of batch levels,
$\bbeta^{(b)} = (\beta_{1}^{(b)}, \dots, \beta_{n_{b}}^{(b)})^{T}$ with $c_{b}$
constraints.  Then the degrees of freedom are $\nu_{b} = n_{b} - c_{b}$, and the
finite-population variance $s^{2}_{b}$ is $s^{2}_{b} = \frac{1}{\nu_{b}}
{\bbeta^{(b)}}^{T} \left( \I_{n_b} - \C_b^T ( \C_b \C_b^T )^{-1} \C_b \right)
\bbeta^{(b)}$, where $\I_{n_b}$ is the $n_b \times n_b$ identity and $\C_b$ is
the $c_b \times n_b$ constraint matrix such that $\C_b \bbeta^{(b)} = \0$.
Variance component estimation is made by decomposing the variance of the batch
level estimates, $V_b = \var ( \widehat{\beta}^{(b)} ) = \var \{ \E (
\widehat{\beta}^{(b)} \mid \beta^{(b)} ) \} + \E \{ \var (
\widehat{\beta}^{(b)}\mid \beta^{(b)} ) \}$, into the sum of the superpopulation
variation, $\sigma^2_b$, plus the variability of the batch level estimations,
$V_{\textrm{b:estimation}}$.
The chosen estimate of the superpopulation variance, in this case the
method-of-moments estimator, is then
$\widehat{\sigma}^{2}_{b} = \widehat{V}_{b} -
\widehat{V}_{\textrm{b:estimation}}$, where, $\widehat{V}_{b} =
\frac{1}{\nu_{b}} \sum_{j=1}^{n_{b}} (\widehat{\beta}_{j}^{(b)})^{2}$, and
$\widehat{V}_{\textrm{b:estimation}} = \sum_{k \in I(b)} \frac{n_{b}}{n_{k}}
\widehat{\sigma}^{2}_{k}$ includes superpopulation variances from other batches
that enter into variability of the batch level estimates, indicated by the set
$I(b)$.  At a minimum, $\widehat{V}_{\textrm{b:estimation}}$ will include
$\widehat{\sigma}^2_B$, the estimated error variance.  For a large class of
multilevel, hierarchical models, this strategy allows for all terms to be
treated as random, and for their variabilities to be assessed.  

\subsubsection{Confirmatory Procedures}\label{subsubsec:infer1}
In regards to more inferential procedures, either a frequentist or Bayesian
direction can be taken.  In the frequentist case, an inverse-chi-square
distribution, $\chi^{-2}_\nu$, is employed to assess uncertainty in the
superpopulation variance $\sigma^2_b$; since $\frac{1}{\nu_{b}}s^{2}_{b} /
\sigma^{2}_{b}$ is chi-square distributed with $\nu_b$ degrees of freedom.  For
batches with $\sum_{k \in I(b)} \frac{n_{b}}{n_{k}} \widehat{\sigma}^{2}_{k}$,
including more than only the error variance $\widehat{\sigma}^2_B$, then a
linear combination of inverse-chi-square distributions, $\sum_i m_i
\chi^{-2}_{\nu_i}$, is required, as described at the end of the previous
section.  As \citet{Gelm:05a} states, these linear combinations may be dealt
with directly, although simulation is often more straightforward.  The
simulation, described therein, is carried out by: 1) Obtain $R$ simulated raw
variances, $V_{b}$ for each of the $B$ batches in the model with a random
variable that is proportional to $\chi^{-2}$ and corresponding degrees of
freedom; 2) Calculate superpopulation variances using $\widehat{\sigma}^{2}_{b}
= \max(0,V_{b} - V_{\textrm{b:estimation}})$; 3) Simulate batch levels using
newly generated superpopulation variances; 4) Calculate sample variances of each
batch.  This procedure then yields a (posterior) sample of superpopulation
variances, batch levels, and finite-population variances, corresponding to the
final three steps.

A strict Bayesian approach requires additional prior specifications, but yields
posterior distributions of the superpopulation variances.  However, this
distinction, between the two schools of thought, can be seen as purely semantic.
As \citet{Gelm:05a} states, ``given $\sigma^2_b$, the parameters $\beta_j^{(b)}$
have a multivariate normal distribution (in Bayesian terms, a conditional
posterior distribution; in classical terms, a predictive distribution)''.  Thus,
assessing uncertainty in the finite-population variances, $s^2_b$, is the same.
In both cases, either batch levels themselves are simulated, or the distribution
of $s^2_b$ is approximated with an appropriate chi-square random variable.

It should be clear that the uncertainty surrounding superpopulation variance
parameters will typically be greater than that for finite-population variances.
Intuitively, this is because superpopulation variances describe variability of
levels that have not yet been realized.

\subsection{Multilevel Multivariate ANOVA}\label{subsec:banova}
Typical multivariate analysis of variance strategies rely on the distribution of
the determinant of sums of squares matrices, i.e.\ Wilks' lambda distribution
\citep[p. 335]{Mard:etal:79}, and culminate in $p$-value related conclusions.
Hence, it does not easily facilitate inclusion of random effects, and thus no
comparison across these effects.  Using a Bayesian approach, we now derive a
general multivariate methodology that seeks to provide results similar to those
of Section~\ref{subsubsec:gelmananova}.  The method partitions variability by
batch in an efficient manner and further factors the posterior by batch into a
batch's superpopulation covariance posterior, and batch levels posteriors.  In
addition, we handle the issue of constraints in a way that does not commit what
\citet{Neld:94} has called one of the false steps of linear models.  Rather, the
constraints are implicit, yielding improper batch level prior distributions.
Another point which must be mentioned is that of matrix parameter estimation.
Although the common limitations of covariance matrix estimation and modeling are
issues that must be dealt with, it is important to first focus on, and refine
multivariate ANOVA for familiar settings.  In Section~\ref{sec:discuss} we
discuss issues, modifications, and implications of the method when higher
dimensional data is used. 

\subsubsection{Multivariate Model Parametrization}\label{subsubseq:multivarhier}
Consider $d$-dimensional multivariate observations such that batch levels are
vectors and batch variances are covariance matrices.  Namely,
\eqref{eq:anovamodel1}--\eqref{eq:randeffeq} now contain vectors $\Y_i$,
$\bbeta_{j^b}^{(b)}$, matrices $\X_{i,j}$, and covariance matrices are
$\bSigma_b$, all of appropriate dimension.  We adopt the strategy from the
previous section, in indiscriminately considering all terms as potentially
possessing variability.  The multivariate analogue of \eqref{eq:anovamodel1}
with Gaussian errors is 

\vspace{-0.55cm}
\begin{align}
  \Y_{i} \mid \{\bbeta_{j^b}^{(b)} \}_b, \bSigma_\epsilon & \sim
    \cN_d \left( \sum_{b = 0}^{B - 1} \bbeta_{j^b_i}^{(b)}, \; \bSigma_\epsilon
    \right), \qquad & i = 1, \dots, n,
    \label{eq:manov1yi} 
\end{align}
and the remainder of the Bayesian model specification is given by

\vspace{-0.55cm}
\begin{align}
  \bbeta_{j}^{(b)} \mid \bSigma_b & \sim \cN_d(\bbeta_0^{(b)}, \; \bSigma_b),
    \qquad & j = 1, \dots, n_b,
    \label{eq:manov1betaj} \\
  \U_b = \bSigma_b + \frac{n_b}{n} \bSigma_\epsilon \mid \bSigma_\epsilon & \sim
    W^{-1}(\bPsi_b, \; \kappa_b), 
    \label{eq:manov1sigb} \\
  \bSigma_\epsilon & \sim W^{-1}(\bPsi_\epsilon, \; \kappa_\epsilon) 
    \label{eq:manov1sigep},
\end{align}
for $b = 0, \dots, B-1$ and with the inverse-Wishart distribution denoted by
$W^{-1}$.  Batch indices $b = 0$ and $b = B$ respectively correspond to the
intercept, or overall mean term, $\bmu$, and to measurement errors,
$\bepsilon_i$.  For notational convenience we will refer to $\bSigma_{\mu}$ and
$\bSigma_{\epsilon}$, rather than $\bSigma_0$ and $\bSigma_B$.  Typically
zero-mean batch level priors are assumed; that is, $\bbeta_0^{(b)} = \0$.
Setting $\bPsi_b = \0$, $\kappa_b = 0$ yields the noninformative prior $p(\U_b)
\propto \vert \U_b \vert^{-(d+1)/2}$.  Because inverse-Wishart support is given
by the set of all positive definite matrices, \eqref{eq:manov1sigb} is referred
to as a constrained inverse-Wishart distribution, since $\U_b - \frac{n_b}{n}
\bSigma_\epsilon$ is required to be positive definite.  This covariance
parametrization, $\U_b = \bSigma_b + \frac{n_b}{n} \bSigma_\epsilon$,  has
previously been utilized in the context of multivariate random effects by
\citet{Ever:Morr:00b}, for which they develop efficient methods of sampling.
Also, when only error terms contribute to the variance of batch level estimates,
$\U_b$ is analogous to $V_b$ of Section~\ref{subsubsec:gelmananova} with $I(b) =
\{ B \}$.  

The choice of inverse-Wishart covariance priors, \eqref{eq:manov1sigb} and
\eqref{eq:manov1sigep}, has been made to balance the complexity of model
specification with implementation and computation.  However, given the
considerable amount of research of covariance priors, there are other options
available.  \citet{Dani:99} and \citet{Dani:Kass:01} examine covariance priors
that emphasize uniform shrinkage of the eigenvalues.   Although informative
covariance priors may be necessary in many cases, the usage of such priors will
impact the computational demands required.  

\subsubsection{Posterior Distributions}\label{subsubsec:infer2}
Without additional specification, \eqref{eq:manov1yi}--\eqref{eq:manov1sigep}
yield an inadequate posterior.  In an MCMC setting this may manifest itself by
failure to converge, due to drifting in the parameter space.  From a classical
point of view, estimating the set of all batch levels, $\{ \bbeta_{j^b}^{(b)}
\}$ would require additional constraints.  The inclusion of similar constraints
in the Bayesian model allows for a closed form of the posterior, as well
as for factorization between batches.  Degrees of freedom for each batch are
then accounted for in the corresponding batch covariance posterior.  This
parametrization is also beneficial in terms of computation since batches are
conditionally independent of one another.  Using a vectorized form of the model,
$\Y = (\Y_{1}^T, \dots,\Y_{n}^T)^T \in \IR^{nd}$ and $\bbeta^{(b)} =
(\bbeta^{(b)^T}_1, \dots, \bbeta^{(b)^T}_{n_b})^T \in \IR^{n_b d}$, is
convenient for the development.  The constraint $\C_b \bbeta^{(b)} = \0 \in
\IR^{c_b d}$, where there are $c_b$ constraints, combined with
\eqref{eq:manov1yi}, \eqref{eq:manov1betaj}, is now

\vspace{-0.55cm}
\begin{align}
  \Y \mid \{\bbeta^{(b)} \}_b, \C_b \bbeta^{(b)} =\0, \, \bSigma_\epsilon & \sim
    \cN_d \left( \sum_{b=0}^{B-1} \bbeta^{(b)}, \; \I_{n} \otimes
    \bSigma_\epsilon \right), 
    \label{eq:manov1ystar}\\
  \bbeta^{(b)} \mid \C_b \bbeta^{(b)} = \0, \, \bSigma_b & \sim \cN_{n_b d}
    ( \1_{n_b} \otimes \bbeta_0^{(b)}, \; \widetilde{\bOmega}_b ), 
    \label{eq:manov1betastar}
\end{align}
with $\otimes$ denoting the Kronecker product.  The rank-deficient
$\widetilde{\bOmega}_b$, due to the constraint, causes \eqref{eq:manov1betastar}
to be improper.  To derive this improper distribution begin with the
unconstrained and vectorized form of $\eqref{eq:manov1betaj}$, which has
covariance $\bOmega_b = \I_{n_b} \otimes \bSigma_b$.  The density is then stated
through a decomposition of the precision, $\Q_b = \bGamma\bLambda\bGamma^T
\otimes \bSigma_b^{-1}$.  Assuming $c_b$ constraints, the rank deficiency is
introduced by removing the corresponding number of eigenvalues from the diagonal matrix
$\bLambda$, e.g.\ $\widetilde{\bLambda} = \textrm{diag}(0, \dots, 0,
\lambda_{c_b + 1}, \dots, \lambda_{n_b})$, leading to $\widetilde{\Q}_b = \bGamma
\widetilde{\bLambda} \bGamma^T \otimes \bSigma_b^{-1}$.  This method of
addressing linear constraints is useful for other general improper distributions
and intrinsic Gaussian Markov random fields, as illustrated by
\cite{Rue:Held:05}.  Let $\vert\Q\vert_{*}$ be the pseudo determinant of a
singular matrix, that is, the product of its non-zero eigenvalues.  In the case
of the identity matrix being used as the first matrix term in the Kronecker
product, the eigenvalues are one, thus densities of the likelihood,
\eqref{eq:manov1ystar}, and of batch level priors, \eqref{eq:manov1betastar}
are 

\vspace{-0.55cm}
\begin{align*}
  p( \Y \mid \{\bbeta^{(b)}_j \}_b, \bSigma_\epsilon) \propto & \vert
    \bSigma_\epsilon \vert^{-n/2} \textrm{exp} \left(
    -\frac{1}{2}\sum_{i=1}^n(\Y_{i} - \widehat{\Y}_{i})^T \bSigma_\epsilon^{-1}
    (\Y_{i} - \widehat{\Y}_{i}) \; \right. \\ 
    & \left. \qquad \qquad \quad + \; -\frac{1}{2}\sum_{b=0}^{B-1} \sum_{j=1}^{n_b} (\bbeta^{(b)}_j
    - \widehat{\bbeta}^{(b)}_j)^T \frac{n}{n_b}\bSigma_\epsilon^{-1}
    (\bbeta^{(b)}_j - \widehat{\bbeta}^{(b)}_j) \right), \\
  p( \bbeta^{(b)} \mid \bSigma_b )
    \propto & \vert \widetilde{\Q}_b \vert_{*}^{1/2} \textrm{exp} \left( -\frac{1}{2}
    ( \bbeta^{(b)} - \1 \otimes \bbeta^{(b)}_0 )^T \widetilde{\Q}_b ( \bbeta^{(b)} -
    \1 \otimes \bbeta^{(b)}_0 ) \right) \\ 
    \propto & \vert \bSigma_b \vert^{-(n_b-c_b)/2} \textrm{exp} \left(
    -\frac{1}{2}\sum_{j=c_b+1}^{n_b} ( \bbeta^{(b)}_j - \bbeta^{(b)}_0 )^T
    \bSigma_b^{-1} ( \bbeta^{(b)}_j - \bbeta^{(b)}_0 ) \right),
\end{align*}
where $\widehat{\,\cdot\,}$ denotes least-squares estimates.  For orthogonal
batches, the full posterior from these densities and from batch covariance
prior densities, can be conveniently factored into

\vspace{-0.55cm}
\begin{align}
  p(\bSigma_\epsilon, \; \{ \bSigma_b, \bbeta^{(b)} \}_{b=0}^{B-1} \mid \Y ) 
    = p(\bSigma_\epsilon \mid \Y ) \prod_{b=0}^{B-1} 
    p(\bSigma_b, \; \bbeta^{(b)} \mid \Y, \; \bSigma_\epsilon). \label{eq:fullposttmp}
\end{align}
Each joint batch density, $p(\bSigma_b, \bbeta^{(b)} \mid \Y,
\bSigma_\epsilon)$, is then factored further.  Using known matrix identities,
e.g.\ $( \A^{-1} + \B^{-1} )^{-1} = \A ( \A + \B )^{-1} \B$, the identity

\vspace{-0.55cm}
\begin{align}
\begin{split} 
  (\x-\s)^T \U^{-1} (\x-\s) + (\x-\t)^T \V^{-1} (\x-\t) & \\ 
  = \; \mathrm{tr}\left( (\U + \V)^{-1} (\s - \t) (\s - \t)^T \right) & \; + 
  \; (\x - \P^{-1}\m)^T \P (\x - \P^{-1}\m),
  \label{eq:matident1}
\end{split}
\end{align}
is derived, where $\P = \U^{-1} + \V^{-1}$, $\m = \U^{-1}\s + \V^{-1}\t$.
Accounting for model constraints, together with \eqref{eq:matident1}, the batch
superpopulation posterior and the batch levels are found through the
decomposition of quadratic forms of batch levels and least squares estimates

\vspace{-0.55cm}
\begin{align*} 
  \mathrm{tr}\left[ (\bSigma_\beta + \frac{n_b}{n} \bSigma_\epsilon)^{-1}
    \B_b \right]
  + \sum_{j=1}^{c_b} (\bbeta^{(b)}_j - \widehat{\bbeta}_j^{(b)})^T \frac{n}{n_b}
    \bSigma_\epsilon^{-1} (\bbeta^{(b)}_j - \widehat{\bbeta}_j^{(b)}) \\ \quad +
    \sum_{j=c_b+1}^{n_b} (\bbeta^{(b)}_j - \P_b^{-1} \m^{(b)}_j)^T \P_b (\bbeta^{(b)}_j -
    \P^{-1}_b \m^{(b)}_j),  
\end{align*}
where $\P_b = \bSigma_b^{-1} + \frac{n}{n_b}\bSigma_\epsilon^{-1}$, $\m^{(b)}_j
= \frac{n}{n_b}\bSigma_\epsilon^{-1} \widehat{\bbeta}^{(b)}_j + \bSigma_b^{-1}
\bbeta_0^{(b)}$, and tr$(\cdot)$ denoting the trace operator.  Additionally,
$\B_b = \sum_{j = c_b + 1}^{n_b} ( \widehat{\bbeta}^{(b)}_j - \bbeta^{(b)}_0 ) (
\widehat{\bbeta}^{(b)}_j - \bbeta^{(b)}_0 )^T$, is analogous to a matrix sums of
squares of the unconstrained batch level estimates that has been adjusted by the
prior mean.  The full joint posterior is then factored as

\vspace{-0.55cm}
\begin{align}
  p(\bSigma_\epsilon, \; \{ \bSigma_b, \, \bbeta^{(b)} \}_{b=0}^{B-1} \mid \Y ) 
    = p(\bSigma_\epsilon \mid \Y ) \prod_{b=0}^{B-1} p(\bSigma_b \mid \Y, \; \bSigma_\epsilon) \;
    p(\bbeta^{(b)} \mid \Y, \; \bSigma_\epsilon, \; \bSigma_b), \label{eq:fullpost}
\end{align}
where the product denotes batch posterior independence, and thus no need for
computationally intensive MCMC procedures.  The corresponding distributions of
\eqref{eq:fullpost} are

\vspace{-0.55cm}
\begin{align}
  \bSigma_\epsilon \mid \Y & \sim W^{-1}\left( \bPsi_\epsilon +
    \sum_{i=1}^n (\Y_{i} - \widehat{\Y}_{i})(\Y_{i} - \widehat{\Y}_{i})^T, \;
    \kappa_\epsilon + n - \sum_{b=0}^{B-1} n_b \right),
    \label{eq:post1sigep} & \\
  \bSigma_b + \frac{n_b}{n}\bSigma_\epsilon \mid \Y, \bSigma_\epsilon & \sim W^{-1} \left(
    \bPsi_b + \B_b,
    \; \kappa_b + n_b - c_b \right), &
    \label{eq:post1sigb} \\
  \bbeta^{(b)}_j \mid \Y, \bSigma_\epsilon, \bSigma_b & \sim
  \begin{cases}
    \cN_d \left( \widehat{\bbeta}^{(b)}_j, \; \frac{n_b}{n}\bSigma_\epsilon
    \right) 
      \qquad & j = 1,\dots, c_b,
      \label{eq:post1beta} \\
    \cN_d \left( \P^{-1}_b \m^{(b)}_j, \; \P^{-1}_b \right) \qquad & j=c_b+1, \dots, n_b.
  \end{cases}  
\end{align}
Batch levels \eqref{eq:post1beta}, which reflect both free and constrained
parameter estimates, can then be sampled and adjusted accordingly to obtain
posteriors for finite-population covariances, $\S_b$.  Recall finite-population
parameters focus on observed levels of a batch, not on all potential unobserved
batch levels.  

\subsubsection{Covariance Posteriors}\label{subsubsec:connec}
In all cases thus far covariances $\bSigma_b$ are assumed to be of full rank.
Hence, improper covariance posteriors will be due only to an insufficient number
of observed levels of the batch.  \citet{Diaz:Garc:97} offer a comprehensive
look at all possible cases of improper Wishart distributions and following their
terminology this would be classified as a pseudo-inverse-Wishart.
\citet{Uhli:94} as well as \citet{Sriv:03} consider sampling with a
pseudo-singular-Wishart distribution.  Extending their work to inverse-Wishart
distributions is one method for dealing with moderate discrepancies in the
number of observed levels, $n_b < d$.  Addressing cases in which $n_b \ll d$ is
discussed in Section~\ref{sec:discuss}.  Even in the remaining case, $n_b \geq
d$, simulation from a posterior is not always efficient.  Because support of the
inverse-Wishart posterior requires positive definiteness in two respects,
$\bSigma_\beta > 0$, and $\bSigma_\beta - \frac{n_b}{n}\bSigma_\epsilon > 0$,
the usual method of rejection sampling from an inverse-Wishart distribution is
not always practical.   \citet{Ever:Morr:00b} describe a more computationally
efficient method to maintain positive definiteness through a Cholesky
decomposition and maintaining positive eigenvalues while the sample realization
is generated.

\subsubsection{Analysis Results}\label{subsubsec:communresults}
Multivariate sources of variability do not always yield a single, clear
criterion that indicates the greatest contributor to overall variability.  For
scalar variance components, $\sigma^2_b > \sigma^2_{\epsilon}$ is clearly
interpreted, however, due to the partial ordering of positive definite matrices,
the analogous statement on covariance matrices is not useful.  In other words,
there is not a single, obvious comparison that can be made to determine which of
two covariances are ``greater''.   Depending on the setting, there may exist an
adequate scalar that sufficiently summarizes covariance characteristics.  For
the volume of ellipsoidal contours the determinant, $\vert\bSigma\vert$,
achieves this, while in other cases the sum of all entries, $\1^T\bSigma\1$, or
the sum of the marginal variances, $\mathrm{tr}(\bSigma)$, may be appropriate.
Scalar criteria with corresponding uncertainty intervals then allow multivariate
sources of variability to be directly compared.  \citet{Mard:etal:79} employed
the determinant and trace, which correspond to the product and sum of
eigenvalues; referring to them as the generalized variance and total variance,
respectively.  We have amended the terminology, since inclusion of the sum of
all matrix elements requires further delineation, by denoting $\1^T\bSigma\1$ as
the total variance, and $\mathrm{tr}(\bSigma)$ as the total marginal variance.

Effectively relaying results of analysis of variance is one of the motivating
factors that \citet{Gelm:05a} cites.  The classical table of $p$-values does not
yield any indication of batches with the largest variances, nor are the required
assumptions on other batches clear.  \citet{Neld:99} discusses many of the
issues of over-reliance upon the $p$-value, and its ineffectiveness as a tool
for communicating results.  Uncertainty intervals are the default choice for
presenting results.  Visual plots are convenient since they facilitate
simultaneous comparison of the relative variability contributions, their
magnitudes, and the magnitude of the uncertainty in the estimates.  For direct
comparison of batches $b$ and $b^{\prime}$, statements of the form $P \bigl( g(
\bSigma_b ) > g( \bSigma_{ b^{\prime} } ) \; \vert \; \Y \bigr)$, utilizing an
arbitrary matrix criterion $g(\cdot)$, may also be found.

\section{Examples}\label{sec:examples}
This section covers two examples of the outlined methodology to carry out
confirmatory procedures on multivariate data.  The first is a toy example in
which output and summaries are given in order to provide further insight.  The
second example utilizes global averages of temperature and precipitation
predictions using $13$ atmosphere-ocean general circulation models and $3$
global greenhouse gas emissions scenarios that have been identified by the
Intergovernmental Panel on Climate Change.

\subsection{Simulation}\label{subsec:simul}
The process $\Y_{ij} = \bmu + \balpha_i + \bepsilon_{ij} \in \IR^d$, i.e.\ $B =
2$, $d = 3$, is used to illustrate the method of Section~\ref{subsec:banova}.
To generate the three-dimensional observations, parameters $\bSigma_\alpha,
\bSigma_\epsilon$, and $\bmu$ are fixed.  An individual simulation is then
performed by generating $\balpha_i$ and $\bepsilon_{ij}, i = 1, \dots, n_\alpha,
\; j = 1, \dots, n_\epsilon$, from mean-zero multivariate Gaussian distributions
with their respective fixed covariances.  The mean term $\bmu$ is added to
generated data resulting in $n = n_\alpha n_\epsilon$ observations.  

Using \eqref{eq:post1sigep}--\eqref{eq:post1beta}, posterior distributions for
these covariance criteria are obtained for three distinct simulation scenarios.
The three scenarios can be explained by the ambiguous description that variability
introduced by batch $\balpha$ is greater than (case 1), less than (case 2), or
comparable to (case 3) variability introduced by error batch $\bepsilon$.  More
specifically, covariance matrices are decomposed into a vector of marginal
standard deviations $\s$ and a correlation matrix $\R$, e.g.\ $\bSigma_\alpha =
\mathrm{diag}(\s_\alpha) \R_\alpha \mathrm{diag}(\s_\alpha)$.  For all
simulation correlation matrices are fixed.  Batch levels $\balpha_i$ have the
unique correlation structure $(\R_{\alpha})_{1,2} = 0.3, (\R_{\alpha})_{1,3} =
0.1, (\R_{\alpha})_{2,3} = 0.5$. Errors $\bepsilon_{ij}$ have the correlation
matrix, $\R_{\epsilon}$, with autocorrelation structure $(\R_{\epsilon})_{i,j} =
\rho^{\vert i - j \vert}$, and $\rho=0.2$.  Error marginal variances are
additionally held constant at $1$ over all simulations, $\bSigma_\epsilon =
\R_\epsilon$, so that only $\s_\alpha$ is distinct for each case.

\begin{figure}[b!]
\centering
\includegraphics[width=\textwidth]{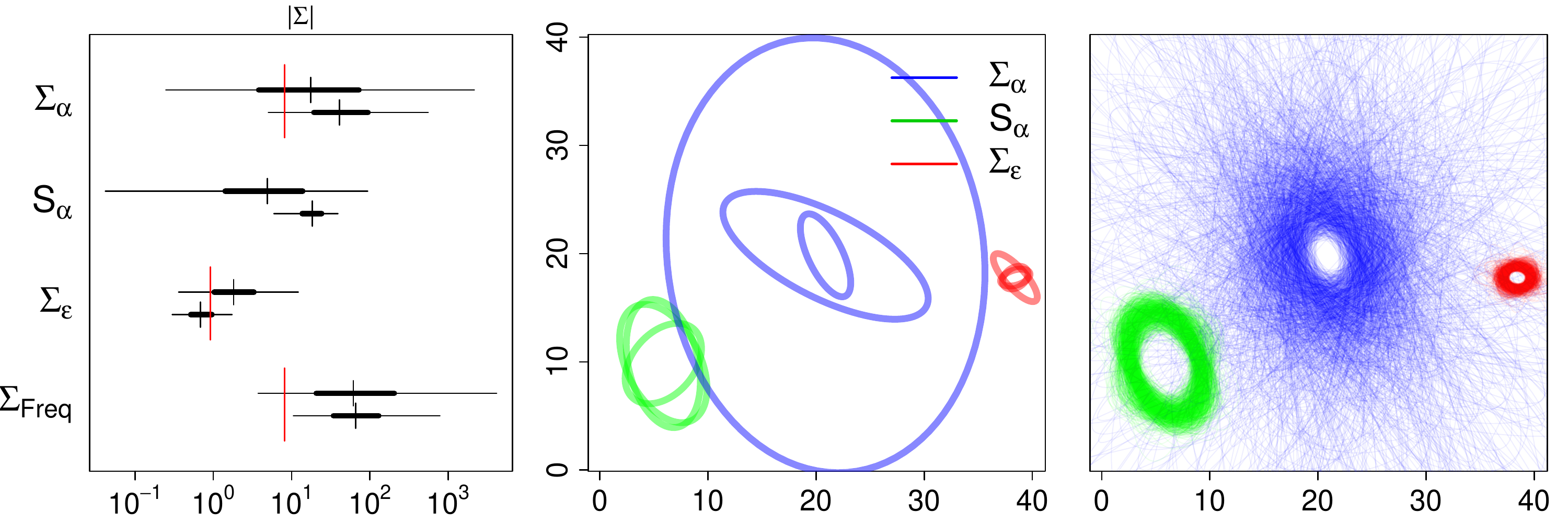}\\[-4mm]
\caption{Simulation for two distinct sample size pairs, $n_\alpha = 5,
n_\epsilon = 3$, and $n_\alpha = 8, n_\epsilon = 5$.  In both, batch $\balpha$
varies greater than the errors, with $\s_\alpha = (\sqrt{2}, \sqrt{2},
\sqrt{3})^T$.  Uncertainty intervals with thin narrow
lines denoting $95\%$ uncertainty level endpoints, thicker middle portion
denoting $50\%$ endpoints, and vertical tick at median. Larger red vertical
ticks denote true parameter values.  Higher-positioned, wider intervals
correspond to smaller batch sample sizes, while lower-positioned, narrower
intervals correspond to larger batch sample sizes (left).  For second set of
sample size pairs, ellipses from first two principal components are displayed
for $2.5\%, 50\%,$ and $97.5\%$ determinant-ordered percentiles of posterior
distributions when $1000$ posterior realizations have been drawn (center).  All
$1000$ posterior ellipses (right).} 
\label{fig:pairofsimulations}
\end{figure}

The objective of the analysis is to assess the relative variability introduced
by batch $\balpha$ and batch $\epsilon$, as well as the uncertainty in the
assessment. Further, this is to be done in an appropriate multivariate context.
Figure~\ref{fig:pairofsimulations} displays the results of two simulation runs
under the first scenario (case 1) using the determinant.  In one simulation, the
number of batch level realizations are $n_\alpha = 5, n_\epsilon = 3$ and in the
second $n_\alpha = 8, n_\epsilon = 5$, which is to say less vs.\ more data.  The
left-most graph of Figure~\ref{fig:pairofsimulations} displays uncertainty
intervals, with narrow lines denoting $95\%$ quantiles, thicker lines $50\%$,
and a vertical tick placed at the median.  The upper set of intervals, which are
intuitively wider, correspond to less data, while the lower set of intervals
correspond to more.  By vertical comparison of the uncertainty intervals, we see
that for $n_\alpha = 5, n_\epsilon = 3$ all intervals overlap, and hence no
distinction can be made between the sources of variability.  For $n_\alpha = 8,
n_\epsilon = 5$ however, there is no overlap of the uncertainty intervals,
suggesting that both superpopulation variability, and the finite-population
variability are greater than error variability.  Additionally,
Figure~\ref{fig:pairofsimulations} offers a diagnostic look at covariance
posteriors.  The center figure displays ellipses from first two principal
components corresponding to $2.5\%, 50\%,$ and $97.5\%$ determinant-ordered
percentiles of the posterior distributions, specific values of which can be seen
on endpoints of the uncertainty intervals.  The right-most graph of
Figure~\ref{fig:pairofsimulations}, which shows all $1000$ ellipses from each
posterior distribution, offers a look at the size, shape, and orientation, for
an overall comparison of the uncertainty in the batch covariances.  Size renders
an idea of the magnitude of the marginal variances.  The disparity in size
between the ellipses corresponding to $\S_{\alpha}$ and $\bSigma_{\epsilon}$
suggest that the marginal variances of the former are greater than those of the
latter.  Through shape, one may glean some insight into batch covariance
dependence.  Lastly, the orientation, or varying orientation, suggests the
uncertainty of the dependence, e.g.\ as the orientation of the ellipses
corresponding to $\bSigma_{\alpha}$ fluctuate greatly, there is not much that
can be said about its dependence structure.

For comparison, consider a frequentist approach to a simplified form of the
problem.  Beginning with $\X_i \sim \cN_d(\0, \bSigma), i = 1, \dots, n$, it is
known that $\U = \sum_i (\X_i - \bar{\X}) (\X_i - \bar{\X})^T \sim W_d(\bSigma,
n - 1)$, from which we derive the distributions

\vspace{-0.55cm}
\begin{align}
  \vert\U\vert & \sim \vert \bSigma \vert \cdot \prod_{i=1}^d \chi^{2}_{n-1-d+i},
    \nonumber \\ 
  \1^T\U\1 & \sim \1^T\bSigma\1 \cdot \chi^{2}_{n-1}, \nonumber \\ 
  \mathrm{tr}(\U) & \sim \sum_{i=1}^d \lambda_{i} \cdot \chi^{2}_{n-1}, \qquad 
    \lambda_1,\dots,\lambda_d = \textrm{eig}(\bSigma). 
    \nonumber 
\end{align}
The first two offer pivots and thus allow for closed form expressions that yield
confidence intervals for values of interest $\vert \bSigma \vert, \1^T \bSigma
\1$.  The confidence interval for $\mathrm{tr}(\bSigma)$ is based on the normal
approximation that matches the first two moments, $\E \{ \mathrm{tr}(\U) \} =
(n-1)\mathrm{tr}(\bSigma)$, and $\var \{ \mathrm{tr}(\U) \} =
2(n-1)\mathrm{tr}(\bSigma^2)$.  This approximation has been chosen in the spirit
of moment matching approximations used by \citet{Imho:61} as applied to
quadratic forms of random vectors.  Note that these confidence intervals assume
that the $\balpha_i$ are directly observed, which is not the case for our
proposed method.  Rather, the classical methods shown are
included only for comparison.  

\begin{figure}[h]
\centering
\includegraphics[width=\textwidth]{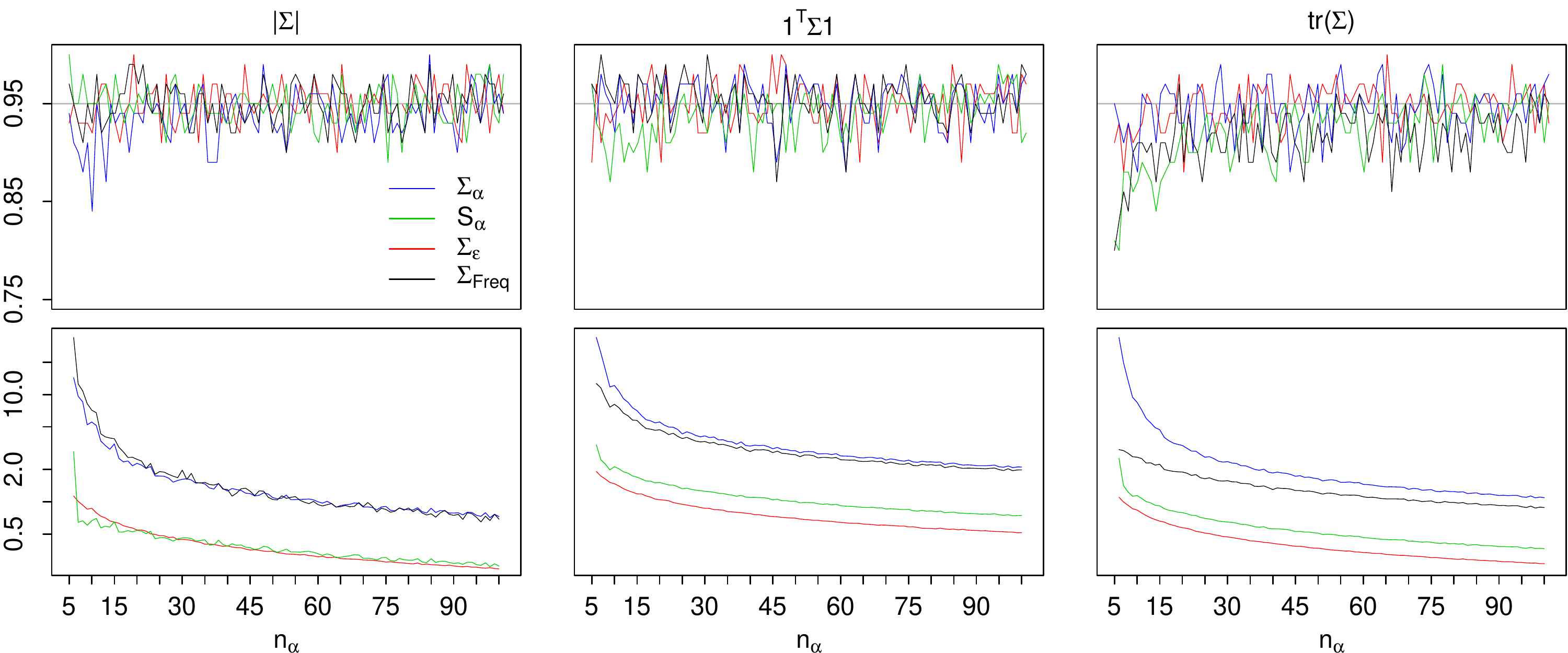}\\[-4mm]
\caption{For $n_\alpha = 5, \dots, 100$, coverage is estimated with $S = 100$
simulations in which the sources of variability are comparable (case 3).  
Coverage estimates for $\bSigma_\alpha$, $\S_\alpha$, and $\bSigma_\epsilon$,
where grey line denotes $95\%$ nominal coverage (top).  Average
uncertainty interval widths on log scale (bottom).} 
\label{fig:coverandwidth}
\end{figure}

To gain insight into the coverage success and uncertainty interval widths, we
have carried out $S=100$ simulations over different values of $n_\alpha$ and
each of the different scenarios of variability sources.  For all simulations the
number of replicates at each level is fixed at $n_\epsilon = 15$.  Results in
all $3$ cases were relatively similar with respect to coverage, noting however
that uncertainty interval widths increase as the magnitude of variability
increases. Thus, only the scenario in which variability sources are comparable
(case 3) has been shown (Figure~\ref{fig:coverandwidth}).  Coverage and interval
widths for $\bSigma_\alpha$ and $\bSigma_{\mathrm{Freq}}$ should be compared as
they correspond to the same true, unknown covariance.  Despite the fact that the
methodology does not assume realizations of $\balpha$ to be observed directly,
but rather indirectly through $\Y$, results are comparable to the frequentist
approach outlined.  

\subsection{Application}\label{subsec:appl1}
In this example our methodology is applied to a bivariate dataset of global
temperature (Celsius) and precipitation (mm/day) for $9$ decadal averages of
boreal summer months, June, July, August, during the remaining century.  The
first batch in the model consists of $13$ levels, each representing a single
atmosphere-ocean general circulation model (AOGCM) developed by several
international climate research institutions as part of the CMIP3 project
\citep{Meeh:00} in the framework of the Fourth Assessment Report (AR4) for the
Intergovernmental Panel on Climate Change (IPCC).  The second batch covers $3$
greenhouse gas emissions scenarios that have been defined by the Special Report
on Emissions Scenarios (SRES), which are identified as A1B, A2, and B1
\citep{Naki:etal:00}.  One fundamental objective of the analysis is then to
compare how these factors contribute to overall variability of global climate
averages, how they relate to one another, and what the uncertainty of this
assessment is.  

Bias and dependence among climate models is an issue that has more recently
begun to be examined further, beginning with \citet{Teba:Knut:07},
\citet{Jun:Knu:Nyc:08}, \citet{Knut:etal:10}, and references therein.  Despite
this, we adopt the statistical assumption that has traditionally been used when
with working with sets of AOGCMs, which is to assume that they are independently
drawn from a common process representative of true climate characteristics.
Using this assumption our approach can be seen as a useful exploratory tool, and
may be further adopted to address contrasts of batch levels, and thus to
identify similar batch levels.  Preliminary analysis steps have suggested the
model 

\vspace{-0.55cm}
\begin{align}\label{eq:gcmmodel}
  \Y_{ijt} = \bmu_0 + \balpha_{0,i} + \bbeta_{0,j} + \bgamma_{ij} +  x_{1,t} \bmu_1 +
  x_{1,t} \balpha_{1,i} + x_{1,t} \bbeta_{1,j} + x_{2,t} \bmu_2 + \bepsilon_{ijt}, 
\end{align}
where $i = 1, \dots, n_\alpha=13$, $j = 1, \dots, n_\beta=3$, $t = 1, \dots,
n_t=9$, $n = n_\alpha n_\beta n_t$, and $d = 2$.  Time covariate $x_{1}$ is
centered such that $x_{1,t} = -4, \dots, 4$, and $x_{2}$ is transformed to
be orthogonal to other predictors in the model.  Batches of interest are AOGCM,
$\balpha$, and SRES, $\bbeta$, and their interaction, $\bgamma$.  The first two
are further specified as a constant effect, $\balpha_0, \bbeta_0$, as well as
with respect to time, $\balpha_1, \bbeta_1$.    

Posterior distributions of batches $\balpha_0, \bbeta_0$, and $\bgamma$ are
derived from \eqref{eq:post1sigb} and \eqref{eq:post1beta}.  Batches $\balpha_1$
and $\bbeta_1$ differ slightly as they correspond to the regression model
formulation.  Multivariate batch levels associated with a covariate would, in
general, be multiplied by a matrix, e.g.\ $\X_{1,t}$.  Using a matrix covariate,
superpopulation and batch level posteriors of batch $\balpha_1$ are then 

\vspace{-0.55cm}
\begin{align}
  \bSigma_{\alpha_1} +  \V_{\alpha_1} \mid \Y, \bSigma_\epsilon & \sim W^{-1} \left(
  \sum_{i=c_{\alpha_1}+1}^{n_\alpha} \widehat{\balpha}_{1,i}
  \widehat{\balpha}_{1,i}^T, \; n_{\alpha} - c_{\alpha_1} \right), &
    \label{eq:post1sigbcovar} \\
  \balpha_{1,i} \mid \Y, \bSigma_\epsilon, \bSigma_{\alpha_1} & \sim
  \begin{cases}
    \cN_d \left( \widehat{\balpha}_{1,i}, \; \frac{n_b}{n}\bSigma_\epsilon
    \right) 
    \qquad & i = 1,\dots, c_{\alpha_1},
    \label{eq:post1betacovar} \\
    \cN_d \left( \P^{-1}_{\alpha_1} \m_{\alpha_1,i}, \; \P^{-1}_{\alpha_1}
    \right) \qquad & i = c_{\alpha_1}+1, \dots, n_{\alpha},
  \end{cases}  
\end{align}
where $\V_{\alpha_1} = \frac{1}{n_\beta} ( \sum_{t=1}^{n_t} \X_{1,t}^T
\bSigma_\epsilon^{-1} \X_{1,t} )^{-1}$, $\P_{\alpha_1} = \bSigma_{\alpha_1}^{-1}
+ \V_{\alpha_1}^{-1}$, and $\m_{\alpha_1,i} = \V_{\alpha_1}^{-1}
\widehat{\balpha}_{1,i}$.  For model \eqref{eq:gcmmodel} the covariate matrix is
$\X_{1,t} = \diag (x_{1,t} )$, and thus $\V_{\alpha_1} = ( n_\beta
\sum_{t=1}^{n_t} x_{1,t}^2 )^{-1} \bSigma_\epsilon$.  The posterior of batch
$\bbeta_1$ is found similarly.

\begin{figure}[h]
\centering
\includegraphics[width=\textwidth]{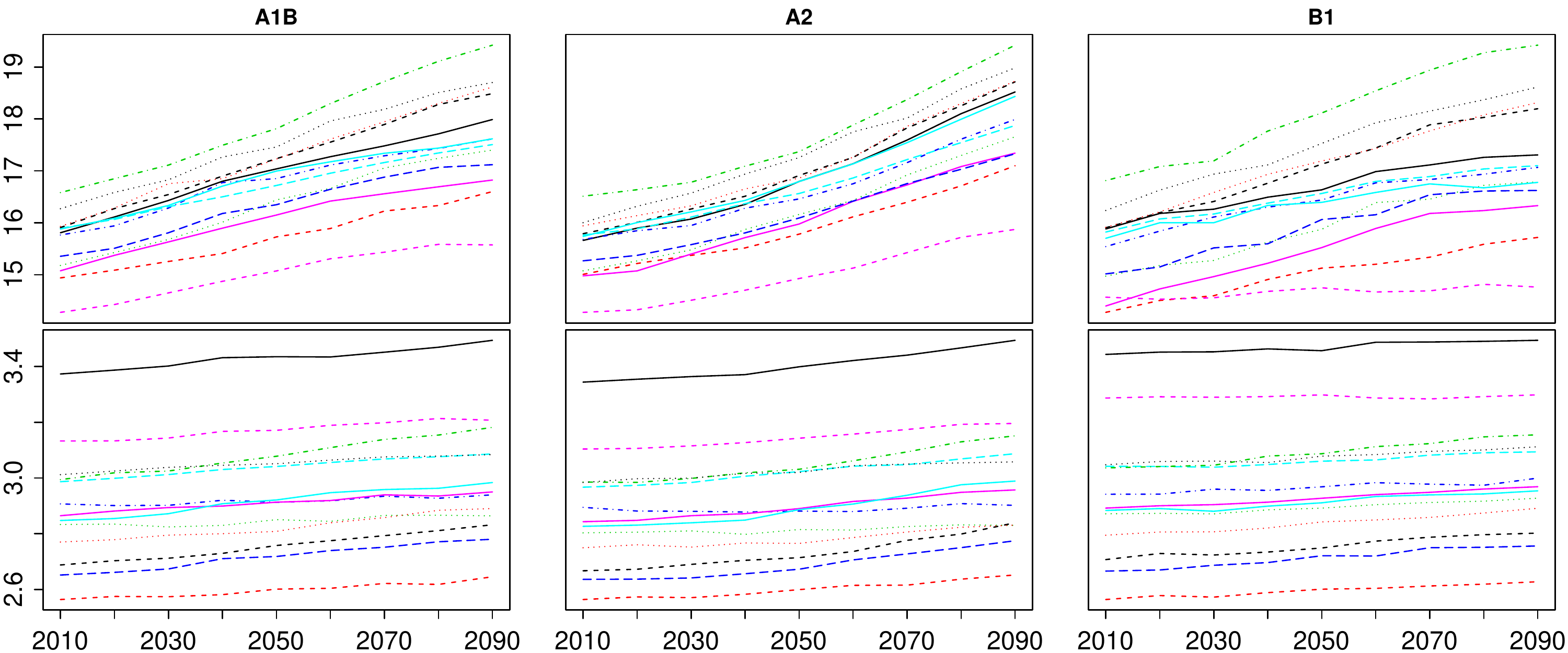}\\[-4mm]
\caption{Global temperature (top) in degrees Celsius and precipitation (bottom)
as mm/day over $n_t=9$ decadal intervals for $n_\alpha=13$ AOGCMs and
$n_\beta=3$ emissions scenarios.} 
\label{fig:gcms1}
\end{figure}

\begin{figure}[h]
\centering
\includegraphics[width=\textwidth]{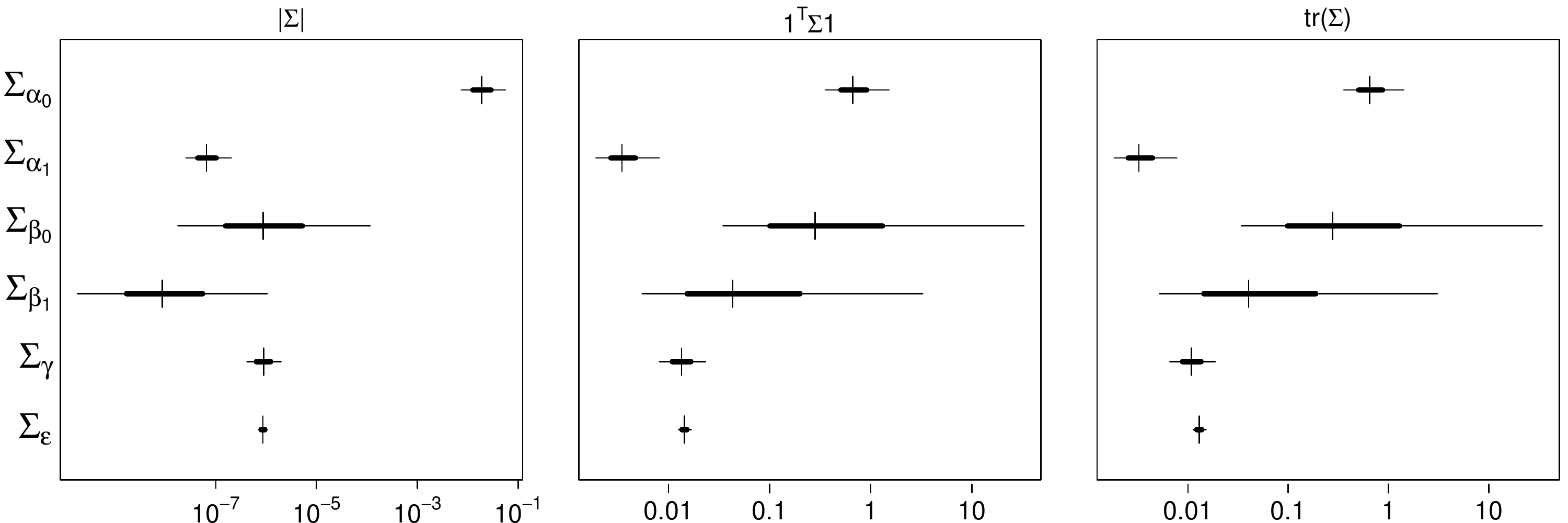}\\[-4mm]
\caption{Superpopulation covariance uncertainty intervals using determinant,
total variance, and total marginal variance criteria shown on a log scale.
Nominal coverages of $0.95$ (thin lines) and $0.50$ (thick lines), and the
median (vertical line) are denoted using quantiles of the corresponding
posterior distributions.}
\label{fig:ex2superpost}
\end{figure}

\begin{figure}[h]
\centering
\includegraphics[width=\textwidth]{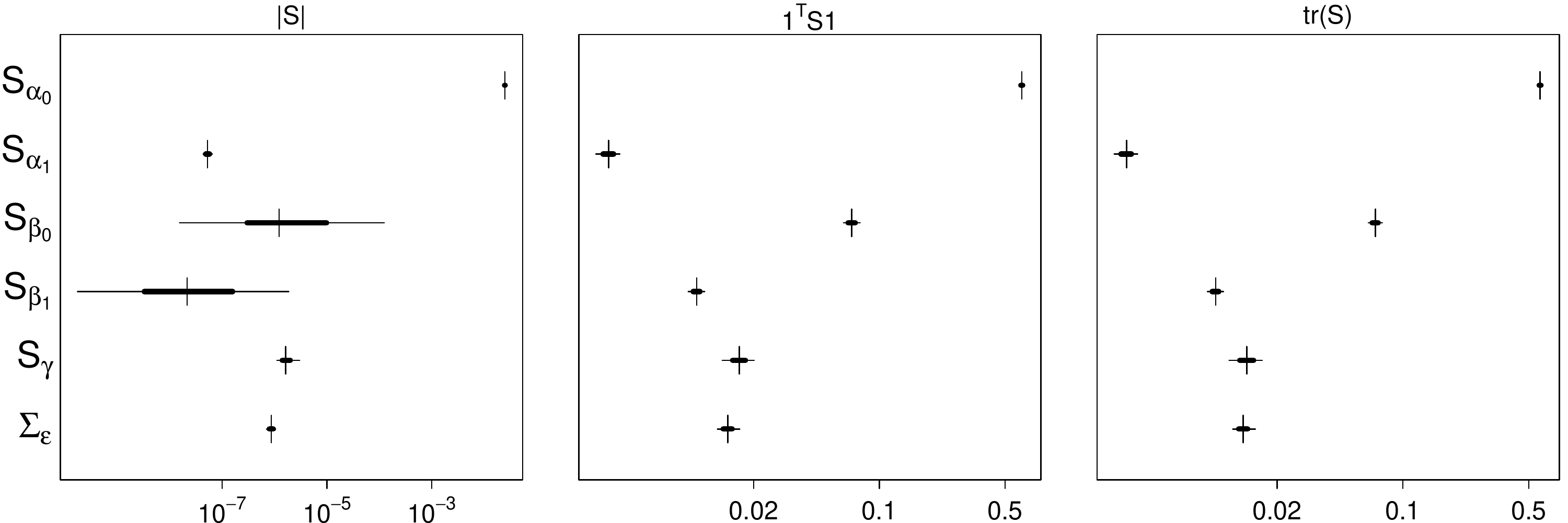}\\[-4mm]
\caption{Finite-population covariance uncertainty intervals using determinant,
total variance, and total marginal variance criteria shown on a log scale.
Nominal coverages of $0.95$ (thin lines) and $0.50$ (thick lines), and the
median (vertical line) are denoted using quantiles of the corresponding
posterior distributions.}
\label{fig:ex2finitepost}
\end{figure}

Figure~\ref{fig:gcms1} suggests that AOGCM is the most distinguishing feature.
Figures~\ref{fig:ex2superpost} and \ref{fig:ex2finitepost} confirm this
assessment since $\balpha_0$ is seen as the most significant source of
variability among all batches.  Comparison of Figures~\ref{fig:ex2superpost},
\ref{fig:ex2finitepost} also illustrate the additional uncertainty of
superpopulation parameters over their finite-population counterparts.
Superpopulation covariance criteria uncertainty intervals are wide because they
account for uncertainty in unobserved batch levels, particularly in the case
when a small number of batch levels have been observed.  Finite-population
covariance uncertainty intervals are generally smaller, because they are
concerned with variability of only the batch levels that have been realized.

\begin{figure}[h]
\centering
\includegraphics[width=\textwidth]{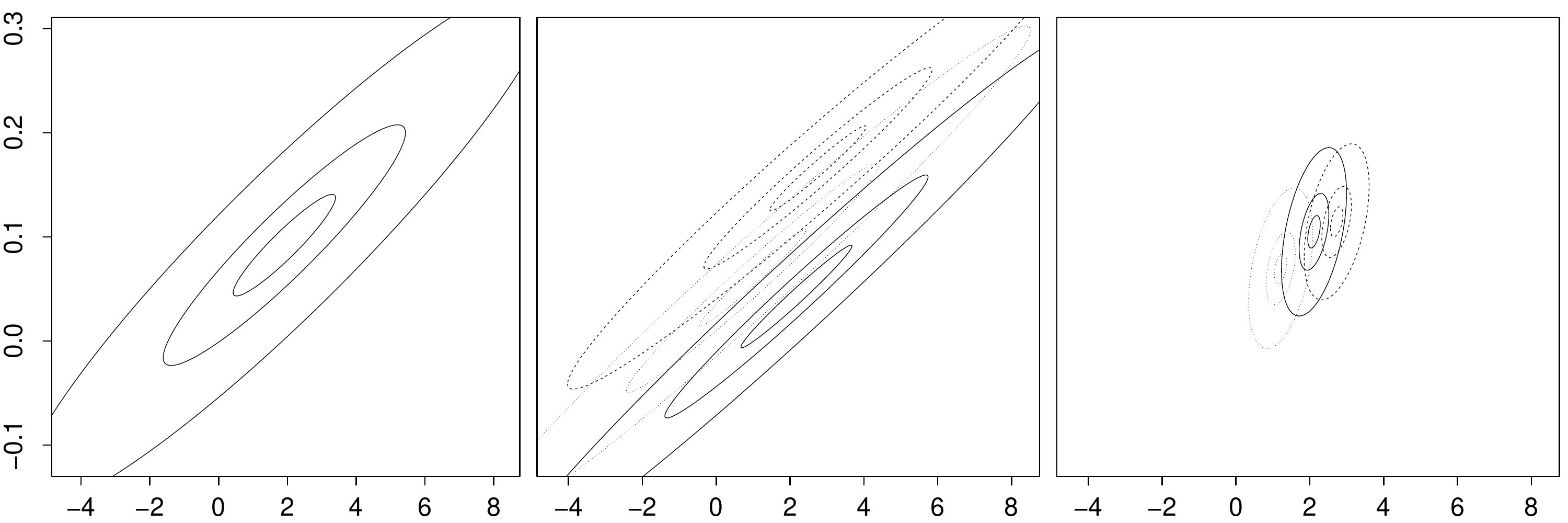}\\[-4mm]
\caption{Posterior predictive distribution, $p( \Y_{i^{\prime} j^{\prime} n_t} -
\Y_{i^{\prime} j^{\prime} 1} \vert \{\Y_{ijt} \} )$, where both $i^{\prime},
j^{\prime}$ represent new, unobserved AOGCM and SRES batch levels (left).
Posterior predictive $p( \Y_{i j^{\prime} n_t} - \Y_{i j^{\prime} 1} \vert
\{\Y_{ijt} \} )$ with observed AOGCM batch levels, $i = 3, 9, 12$
(solid, dashed, dotted) and unobserved SRES batch level
$\bbeta_{1,j^{\prime}}$ (center).  Posterior predictive $p( \Y_{i^{\prime} j
n_t} - \Y_{i^{\prime} j 1} \vert \{\Y_{ijt} \} )$ with unobserved AOGCM batch
level $\balpha_{1,i^{\prime}}$ and all SRES levels, $j = 1,2,3$
(solid, dashed, dotted), that have been observed (right).  Density
contours correspond to quantiles $0.05, 0.25, 0.75$.  Horizontal and vertical
axes denote precipitation in mm/day and temperature in degrees Celsius,
respectively.} \label{fig:ex2postpred}
\end{figure}

Posterior predictive distributions, often used to perform model checking and
diagnostics, can also be utilized to identify distinct sources of variability.
The posterior predictive is conditional on observations with levels from each
batch assumed to be, a) the same as those batch levels that have been observed
data, b) unobserved/novel batch level realizations.
Figure~\ref{fig:ex2postpred} examines posterior density $p(
\widetilde{\Y}_{ijn_t} - \widetilde{\Y}_{ij1} \vert \{ \Y_{ijt} \} )$, the
difference between posterior predictive distributions at the final, $t = n_t$,
and initial, $t = 1$, decades.  Thus, the focus is on temporal batches,
$\balpha_{1i^{\prime}}$, $\bbeta_{1j^{\prime}}$.  Indices $i^{\prime},
j^{\prime}$ signify new, unobserved batch levels.  Linear and quadratic terms,
$\bmu_1, \bmu_2$ are included, although additional variability from these terms
has been disregarded.  The left-most panel, in which every batch contributes a
new batch level realization, shows a large degree of variation.  The center
panel assumes that the AOGCM observed in the original data is to be used, thus
variability from these specific batch level posteriors is included.  For SRES a
novel batch level is assumed, thus a realization utilizing the SRES
superpopulation posteriors is included.  A subset of three observed AOGCM levels
has been displayed, selected so as to best represent the range and relative
distances of their peaks.  However, the high degree of variation introduced by
the new emissions scenario level makes even these distributions nearly
indistinguishable.  In the right-most plot, using observed emissions batch
levels, $n_{\beta} = 3$, the additional variability introduced comes primarily
from the new AOGCM batch level to be observed.  The posterior predictive plots
are particularly useful for determining whether the variability from each batch
is due to the magnitude of the batch variability itself, or due to uncertainty
in the assessment itself. 

\section{Discussion}\label{sec:discuss}
The first contribution of this paper has been in extending recent philosophical
shifts in the treatment of analysis of variance to multivariate settings.  New
analysis of variance approaches allow appropriate parameters, e.g.\ super or
finite population, to be used to answer the correct research question, while at
the same time providing coherent model definition, implementation, and
interpretation.  This same flexibility has been extended to multivariate cases;
in that the researcher can guide covariance criteria choices, rather than the
method determining the criterion.  The second contribution has been in providing
a foundation for computational efficiency, which is necessary for dimension
scalability.  Using improper batch level priors we have shown that it is
possible to minimize dependencies between batch covariances.  In many cases this
reduces, or eliminates, the need for complex and computationally demanding
analyses.




Further extensions to the methodology must explicitly address increasing
dimensionality.  For moderately sized dimensions $d$, relative to number of
observations, improper inverse-Wishart distributions, and/or priors that impose
particular dependence structures, are possible options.  For cases in which $d$
is very large, stricter covariance assumptions may be employed.  In the spatial
context, properties such as stationarity allow covariance parameter space to be
reduced, e.g.\ range, sill, and nugget in a spatial covariance function.
Because simultaneous estimation of such parameters is nontrivial, some
parameters are often assumed, or estimated empirically in earlier analysis
steps, as in \citet{Furr:Sain:Nych:Meeh:07}.  In other cases, so as to maintain
computational feasibility, sparsity restrictions are placed on covariances
\citep{Cres:Joha:08, Furr:Gent:Nych:06, Stei:08}.  For many such scenarios a
covariance is decomposed into a correlation matrix and a scalar variance
parameter.  Our method is then carried out with the inverse-Wishart posterior
density transformed through a spectral decomposition of the correlation matrix,
thus allowing for efficient posterior sampling for cases in which $d \gg n$.
This extension offers an alternative to geostatistical model analyses that have
previously relied on computationally intensive MCMC methods, and is the focus of
current research.  Other difficulties encountered are unbalanced designs and
linearly dependent predictors.  MCMC may be utilized for sets of dependent batch
levels. Development for these cases is another area of current research.
\bigskip

\noindent
\textbf{Acknowledgments} \\
\noindent
We acknowledge the modeling groups, the Program for Climate Model Diagnosis and
Intercomparison (PCMDI) and the WCRP's Working Group on Coupled Modeling (WGCM)
for their roles in making available the WCRP CMIP3 multi-model dataset. Support
of this dataset is provided by the Office of Science, U.S. Department of Energy.

\bibliographystyle{mywiley}
\bibliography{thesis_bib}

\end{document}